\documentstyle[12pt]{article}
\newcommand{\beq}{\begin{equation}}
\newcommand{\eeq}{\end{equation}}
\newcommand{\beqn}{\begin{eqnarray}}
\newcommand{\eeqn}{\end{eqnarray}}

\def\vdir{v\kern-7.8pt\Big{/}}
\def\pdir{p\kern-7.8pt\Big{/}}

\begin{document}
\title{Parton densities beyond Perturbation Theory}
\author{$~~~$}
\date{}
\maketitle

\vskip 0.7truecm

\centerline{ \it U.~Aglietti}

\noindent
Caltech, Pasadena CA, and INFN, Rome

\vskip 0.7truecm

\vskip 0.7truecm

\vskip 0.7truecm

\vskip 0.7truecm

\bigskip
\begin{abstract}

\noindent
I compute non-perturbative corrections
to the kernel governing the evolution
of non-singlet parton densities.
The model used is QED in the limit
of many charged particles.
I find an ultraviolet renormalon corresponding
to a term of order $Q^2/\Lambda^2$, where $\Lambda$
is the pole of the coupling constant.
This term has a non-trivial dependence
on the variable $x=-q^2/(2pq)$ and its coefficient scales as $x^3/(1-x)^2$
($p$ is the momentum of the hadron and $q$ is the momentum transfer).
An extrapolation of my results to QCD implies a breakdown of the
parton model near the elastic region.

\end{abstract}
\bigskip
\bigskip

\bigskip

\noindent
$~^*$ Work supported in part by the U.S. Dept. of Energy under
      Grant No. DE-FG03-92-ER40701.
\newpage

\section{Introduction}

One of the main processes for the study of the hadron structure
is deep inelastic scattering:
\beq\label{eq:proc}
l~+~H~\rightarrow~l~+~X
\eeq
where $H$ is a hadron, $l$ is a lepton and $X$ is any hadronic final state.

\noindent
In the parton model picture, the hadron is viewed as a gas of
fast moving, real, collinear and
non-interacting partons (quarks and gluons),
which share the energy-momentum of the hadron.
Perturbative QCD accounts for hard gluon radiation and
produces logarithmic corrections to the naive parton model.
The cross section $\sigma$ can be written
as the product of a short-distance ($'$hard$'$) cross section times a
parton density:
\beq\label{eq:factoriz}
\sigma(p,q)~=~\sigma_{hard}(xp,q)\cdot q(x, Q^2)
\eeq
where $Q^2=-q^2>0$.

\noindent
In perturbative QCD one can compute the hard cross section
and the evolution of the parton densities from a scale $Q^2$ to
a different scale $Q'^2$ \cite{bpq, books}.
The parton densities at a given scale
cannot be predicted and have to be determined experimentally.

There are corrections to the parton model which originate
from the intrinsic transverse momenta of the partons, their
virtualities,

as well as from parton-parton interactions.
These phenomena are all characterized by the hadronic scale
$\Lambda_{QCD}\sim 300$ MeV. A kinematical analysis shows
that taking these effects into account produces power-suppressed corrections
to the factorized cross section (\ref{eq:factoriz}), of the form
\cite{politz}
\beq\label{eq:nonper}
\left(\frac{\Lambda_{QCD}^2}{Q^2}\right)^n,~~~~~{\rm where}~~~n=1,2,3,...,k,...
\eeq
It is crucial to have an estimate of these non-perturbative terms
for a trustworth application of perturbative QCD.

The occurence of power suppressed corrections can also be demonstrated
with the operator product expansion (OPE) \cite{ope, ope2}.
The cross-section of the process (\ref{eq:proc}) is proportional
to the imaginary part of the forward scattering amplitude
\beq
\langle H \mid h_{\mu\nu} \mid H\rangle
\eeq
where
\beq
h_{\mu\nu}~=~i\int {\rm d}^4 x e^{-iq\cdot x}~
{\rm T}~J_{\mu}(x)J_{\nu}^{\dagger}(0)
\eeq
is a hadronic tensor and $J_{\mu}$ is an
electromagnetic or a weak current.

\noindent
The tensor $h_{\mu\nu}$ can be expanded in the limit,
\beq
Q^2,~p\cdot q~\rightarrow~\infty,~~~~~x={\rm constant},
\eeq
in a tower of covariants of  progressively higher twist $T$
v($T=$ dimension of the operator - spin):
\beq
h_{\mu\nu}~=~\sum_i C_i(Q^2/\mu^2)O_i(\mu^2)
\eeq
where $C_i(Q^2/\mu^2)$ are coefficient functions, $O_i(\mu^2)$
are local operators and $\mu$ is a subtraction point.

\noindent
The lowest twist operators ($T=2$)
\beq\label{eq:lambda}
\overline{\psi}\gamma_{\mu}\lambda_aD_{\nu}D_{\mu_1}D_{\mu_2}...D_{\mu_n}\psi~
-~{\rm traces},~~~~
F_{\mu\rho}D_{\mu_1}D_{\mu_2}...D_{\mu_n}F^{\rho}_{\nu}~-~{\rm traces}
\eeq
correspond to the parton model, while higher twist
operators produce corrections of order $1/Q^2$
($\lambda_a$ is a matrix in flavor space).

Arguments based on parton kinematics or the
OPE can provide only order of
magnitude estimates of the terms (\ref{eq:nonper}).
It is the purpose of this paper to give a semiquantitative
estimate of the power suppressed corrections, based on the
renormalon calculus.
In general, there are power suppressed corrections
to the evolution kernel and to the hard cross section and
the observable effect is a convolution of both terms.
We work hovewer in the leading logarithmic approximation
(see eq.~(\ref{eq:leadlog})),
in which the power suppressed corrections to the
hard cross section are absent. The power suppressed corrections
to the evolution kernel are instead non-vanishing and have a direct
physical meaning.

The idea of the renormalon approach
is that the complete perturbative series
reflects non-perturbative properties of the process \cite{renorm}.
The chain of arguments is the following.
The perturbative series is strongly divergent in all the known cases,
i.e. it diverges for every non-zero value of the coupling constant.
The coefficients have typically a factorial growth,
\beq
\mid c_n~\mid~ \sim n!.
\eeq
A meaning can be assigned to the perturbative series through a resummation
of the series itself.
This operation requires in general an arbitrary prescription, which
introduces an ambiguity. The latter is a
size of the intrinsic non-perturbative corrections.

\noindent
Let us present a physical argument of why
the investigation of the power suppressed corrections (\ref{eq:nonper})
requires the knowledge of the complete perturbative series.
Consider a quantity that has a simple expansion in
powers of $\alpha_s(Q^2)$ (i.e. an infrared safe quantity).
Since
\beq
\alpha_s(Q^2)~\sim~\frac{1}{\log Q^2}~~~~~~{\rm for}~~Q^2\rightarrow\infty,
\eeq
any power of $\alpha_s(Q^2)$ decays more slowly than any power
of $1/Q^2$:
\beq
\left(\frac{1}{Q^2}\right)^m~\ll~\frac{1}{\log^n Q^2}~~~~~~~
{\rm for~any~positive}~n~{\rm and}~m
\eeq
Therefore, it is consistent to look at the power suppressed corrections
only after the inclusion of all the perturbative terms.
The renormalon calculus cannot be applied to QCD because we are not able
to compute the coefficients of any observable for any $n$.
The use of models cannot be avoided and various kinds of $1/N$
expansions have been used. We consider a $1/N$ expansion
of QED, where $N$ is the number of charged particles \cite{qedn}.
This model
is indeed solvable and involves tree diagrams containing strings of
self-enegy corrections to the photon lines (see fig.1).
Since QED contains massless vector quanta like QCD,
collinear singularities are the same in both theories.
Therefore, perturbative physics is the same as in QCD as far as charge flow
effects are neglected.

Actually, QCD can be modelled by changing the sign of the
beta function at the end of our QED-like computation \cite{ben}.
In such a way the infrared properties of QCD are characterized by
the perturbative growth of the coupling constant at small
momentum transfer.
Loosely speaking, we are dealing with tree diagrams
in which the bare coupling is replaced by the QCD one-loop running
coupling constant.

The paper is organized as follows. Section (\ref{bor})
contains a short review of the Borel transform technique and of the
renormalon calculus. In section (\ref{input}) we compute the
input diagrams for the kernel $P_{qq}$.
The latter is derived in section (\ref{ap}).
Section (\ref{disc}) contains a discussion of the results
and section (\ref{concl}) traces the conclusions.
There are also two  appendices. In appendix A the technique
to compute the integrals of the Borel transform is described,
while in appendix B we prove some properties of the planar gauge.

\section{The Borel transform}
\label{bor}

The perturbative expansion of an observable $O$
in quantum field theory is given by a series
of the form
\beq\label{eq:ser}
O(\alpha)~=~c+\sum_{n=0}^{\infty}c_n \alpha^{n+1}.
\eeq
This series is strongly divergent in all known cases because
the coefficients have a factorial growth with $n$: $\mid c_n \mid \sim n!$.

\noindent
As a 0-dimensional model of quantum field theory, we can consider
the integral:
\beq
I_k(\alpha)~=~\int_{-\infty}^{+\infty} {\rm d}x~x^k e^{-x^2-\alpha x^4}
\eeq
There is an instability for $\alpha<0$  and
the expansion in powers of $\alpha$ is  divergent.

By explicit computation one finds
$c_n=(-1)^{n+1}\Gamma(2n+k/2+5/2)/\Gamma(n+2)$.
For large $n$ there is a factorial-like growth of the coefficients
with alternating sign:
$c_n\simeq (-1)^{n+1}2^{2n+k/2+3/2}e^{(k-1)/2}(n+1)!$.

The simplest interpretation of the expansion (\ref{eq:ser}) is that of an
asymptotic series. It is sensitive to truncate the series at its smallest
term, which is a measure of the error.
A more detailed analysis aims to reconstruct
the original function from the knowledge of the $c_n$'s.
This is achieved by means of the Borel transform technique.
The Borel transform is defined by:
\beq\label{eq:borel}
\tilde{O}(t)~=~c~\delta(t) +\sum_{n=0}^{\infty}\frac{c_n}{n!}~t^n
\eeq
We assume that the series on the right-hand side defines an analytic
function of $t$ in a neighborhood of $t=0$, which can be analitically
continued in all the Borel plane.
The observable $O$ can be reconstructed from the Borel trasform
by means of the formula:
\beq\label{eq:invers}
O^B(\alpha)~=~\int_0^{\infty} {\rm d}t~ e^{ -t/\alpha }~\tilde{O}(t)
\eeq
If, for example, $c_n=(-1)^n n!$, the series (\ref{eq:ser}) has zero
radius of convergence, while the Borel transform defines an analytic
function in all the $t$-plane, given by $\tilde{O}(t)=1/(1+t)$.
Since $\tilde{O}(t)$ has no singularities for $t\geq 0$, the observable can be
reconstructed unambigously
by means of the inversion formula (\ref{eq:invers}).
In such a case we say that the perturbative series is Borel summable.
On the contrary, if $c_n=n!$, the Borel transform
is given by $\tilde{O}(t)=1/(1-t)$ and there is a singularity
at $t=1$. The integral in eq.~(\ref{eq:invers}) must be supplemented
with a prescription to deal with the pole.
The arbitrareness of the prescription reflects the intrinsic
ambiguity in the reconstruction of $O(\alpha)$.

Let us discuss now how renormalons are related to power suppressed
corrections in quantum field theory.
Consider a Borel transform of the form:
\beq
\tilde{O}(t)~=~\frac{K}{t-n}~+~{\rm regular~function.}
\eeq
The ambiguity in the observable is given by:
\beq
\delta O~=~K~e^{-n/\alpha(Q^2)}
\eeq
where we have taken the pole residue as a measure of the ambiguity.

\noindent
Expressing the coupling constant in terms of the relevant scales
(dimensional transmutation),
\beq\label{eq:oneloop}
\alpha(Q^2)~=~\frac{1}{-b_0\log Q^2/\Lambda^2},
\eeq
we have:
\beq
\delta O~=~K~\left(\frac{\Lambda^2}{Q^2}\right)^{-b_0 n},
\eeq
where $b_0$ is the first coefficient of the $\beta$-function,
defined including the factor $1/(4\pi)$. In QED, $b_0=1/(3\pi)$.

\noindent
The position of the renormalon pole determines the power
of the non-perturbative correction, while the pole residue sets
the coefficient of the correction.
In general, the Borel transform has a collection of singularities
along the real axis and the leading ambiguity is related
to the pole on the positive axis closest to the origin.

\noindent
Up to now, the existence of renormalons has not been proved in any
realistic theory (such as QED, QCD, $\lambda\phi^4$, etc.).
We consider $QED$ at order $1/N$.
In this model the existence of renormalons is related to the
Landau pole, as shown by the following intuitive argument.
The amplitudes are given by loop  integrals containing
the one-loop running coupling constant (\ref{eq:oneloop})
\beq
\int {\rm d}k^2~\alpha(k^2)~\frac{1}{k^2}
     \frac{1}{(k+p_1)^2}\frac{1}{(k+p_2)^2}\ldots
\eeq
where $p_1,p_2\ldots$ denote external momenta.
The coupling constant diverges at the Landau pole, producing a
divergence of the integral. The latter has to be regularized,
and this operation introduces ambiguities in the amplitude.
There is also a more explicit way of understanding the relation
between Landau pole and renormalons. Expanding the running coupling
constant in powers of $\alpha$ at a fixed scale $\mu$,
\beqn\nonumber
&&\alpha(k^2)~=~\frac{\alpha(\mu^2)}{1-b_0\alpha(\mu^2)\log(k^2/\mu^2)}~=~
\\
&&~~~~~\sum_{n=0}^{\infty}~(b_0\log\frac{k^2}{\mu^2})^n \alpha(\mu^2)^{n+1},
\eeqn
gives rise to a coefficient of $\alpha(\mu^2)^{n+1}$ of the form
\beq
\int_0^1{\rm d}x~\left(\log \frac{1}{x}\right)^n~=~n!
\eeq
There is a fixed sign factorial divergence resulting in renormalons
in the Borel plane.

\section{The computation}
\label{input}

In this section we evaluate the input diagrams for the derivation
of the evolution kernel $P_{qq}$.
The model is $QED$ with $N$ flavors, and we work at order $1/N$.
Quarks are assumed massless and the incoming quark is taken on-shell,
i.e. $p^2=0$. Soft singularities (infrared and collinear)
are regulated by the Borel variable $t$ itself.
For $t>0$ the gluon propagator is indeed less singular than in the
usual case (see appendix A for a formal proof of the regulator).
It is convenient to use the planar gauge with the gauge vector
directed along the final quark direction.
In this gauge collinear singularities decouple from the final quark
leg and the only relevant diagram are
the ladder ones (see appendix B for a proof).
The free photon propagator is given by:
\beq
i\Delta_{\mu\nu}(k)~=~\frac{i S_{\mu\nu}(k)}{k^2+i\epsilon}
\eeq
where
\beq\label{eq:sumpol}
S_{\mu\nu}(k)~
=~-g_{\mu\nu}+\frac{ k_{\mu}n_{\nu}+k_{\nu}n_{\mu} }{ n\cdot k},
\eeq
is the polarization sum, and $n=xp+q$ is a light-like vector, $n^2=0$.

The lowest order diagram is given in fig.2, and is of order $(1/N)^0$.
The rate  is given by:
\beq\label{eq:lo}
w_0~=~\frac{\pi}{2E(p)Q^2}\delta(1-x)
      {\rm Tr}[x\hat{p}\gamma_{\nu}(x\hat{p}+\hat{q})\gamma_{\mu}]
\eeq
where $E(p)$ is the energy of the incoming quark.

\noindent
The next order diagrams involve a single photon exchange, and are
therefore of order $\alpha=a/N$. Inserting bubbles
into the photon line involves the coupling $a=\alpha N$.
Diagrams with multiple gluon emissions are of order
$\alpha^k=a^k/N^k$ with $k\geq 2$.
We work at order $1/N$ and to all orders in $a$. This implies
that we have to consider the diagram of fig.3, with
the photon propagators containing strings of an arbitrary number
of bubbles.

\noindent
The proper self-energy at one-loop is given by:
\beq
i\Pi_{\mu\nu}(k)~=~i(k_{\mu}k_{\nu}-g_{\mu\nu}k^2)\Pi(k^2)
\eeq
where
\beq
\Pi(k^2)~=~-a b_0\log\left(\frac{-k^2-i\epsilon}{\mu^2}\right)
\eeq
in the momentum subtraction scheme.

\noindent
The full photon propagator is therefore given by:
\beqn\nonumber
i \Delta_{\mu\nu}(k)&=&
\frac{i S_{\mu\nu}(k)}{k^2+i\epsilon}~+~
\frac{i S_{\mu\rho}(k)}{k^2+i\epsilon}~i\Pi^{\rho\sigma}(k)~
\frac{i S_{\sigma\nu}(k)}{k^2+i\epsilon} ~+~ ...
\\
&=& \frac{i S_{\mu\nu}(k)}{[k^2+i\epsilon]~[1+\Pi(k^2)]}
\eeqn
The inclusion of the bubbles modifies the free propagator
by the factor $1/[1+\Pi((p-k)^2)]$.

The diagrams involved at order $1/N$ are given in fig.4.
For diagram $(a)$ we have:
\beq
M_a~=~e~\overline{u}(k+q)\gamma_{\mu}\frac{\hat{k}}{k^2+i\epsilon}
\gamma_{\rho}u(p)
\frac{ S^{\rho\nu}(p-k) }{1+\Pi((p-k)^2)}~\epsilon_{\nu}(p-k)
\eeq
The contribution of this diagram to the transition probability per unit time
is given by:
\beqn\nonumber
w_a&=&\frac{1}{4\pi N E(p)}
\int d^4 k~
Tr[\hat{k}\gamma_{\rho}\hat{p}\gamma_{\sigma}\hat{k}\gamma_{\nu}
   (\hat{k}+\hat{q})\gamma_{\mu}]~\times
\\ \label{eq:wa}
&\times&f(a)~
 \delta^+((k+q)^2)~
\delta^+((p-k)^2)
S^{\rho\sigma}(p-k)~~~~~
\eeqn
where $\delta^+(x^2)= \delta(x^2)\theta(x_0)$ and we defined
\beq
f(a)~=~\frac{a}{\mid 1+\Pi(~(p-k)^2)\mid ^2}.
\eeq
The dependence on $a$ is contained in this function, which can be expanded
as
\beq
f(a)~=~\sum_{n,k=0}^{\infty}\frac{(2n+k+1)!}{(2n+1)!~k!}
(-b_0^2\pi^2)^n\Bigg(b_0\log\frac{(p-k)^2}{\mu^2}\Bigg)^k a^{2n+k+1}.
\eeq
According to the definition in eq.~(\ref{eq:borel}),
the Borel trasform is given by:
\beq\label{eq:borelf}
\tilde{f}(t)=
\Bigg(\cos \pi u -\frac{\sin \pi u}{\pi} \log\frac{(p-k)^2}{\mu^2}\Bigg)
\left(\frac{\mu^2}{(p-k)^2}\right)^u
\eeq
where $u=-b_0 t$ (notice a difference of sign with respect
to the usual definition).

\noindent
The Borel transform $\tilde{w}_a$ is computed substituting $\tilde{f}(t)$
in eq.~(\ref{eq:wa}) in the place of $f(a)$.
Because of the $\delta^+((p-k)^2)$, $\tilde{w}_a$ vanishes for $u<0$.
By analytic continuation (we assume that the Borel transform is an analytic
function), the Borel tranform vanishes in all the Borel plane:
\beq\label{eq:zero}
\tilde{w}_a~=~0
\eeq
This result is somewhat paradoxical because it implies the
vanishing of the contribution of the basic ladder diagram,
which is known to produce the one-loop scaling violations
in QCD.
There exists however a physical explanation to eq.~(\ref{eq:zero}).
The dynamics in our model is controlled
by the distribution of invariant masses $(p-k)^2$
flowing into the fermion bubbles. In diagram $(a)$, we have that
$(p-k)^2=0$, implying that any non-trivial contribution
to the Borel transform is impossible.

\noindent
The amplitude of diagram $(b)$ is given by:
\beqn\nonumber
M_b&=&
e^2~\overline{u}(k+q)\gamma_{\mu}\frac{\hat{k}}{k^2+i\epsilon}\gamma_{\rho}
u(p) \frac{S_{\rho\sigma}(p-k)}{[(p-k)^2+i\epsilon][1+\Pi((p-k)^2)]}
\\
&&~~~~~~~~~~~~~~~~~~~~~\overline{u}(p-k-l)\gamma_{\sigma}v(l)
\eeqn
The partial rate is given by:
\beqn\nonumber
w_b&=&\frac{1}{12\pi^2 N E(p)}
\int d^4k~\frac{S^{\rho\sigma}(p-k)}{(p-k)^2}~g(a)~\times
\\ \label{eq:wb}
&\times&\frac{1}{(k^2)^2}~
{\rm Tr}
[(\hat{k}+\hat{q})\gamma_{\mu}\hat{k}\gamma_{\rho}\hat{p}\gamma_{\sigma}\hat{k}
\gamma_{\nu}]~
\delta^+((k+q)^2)~\theta^+((p-k)^2)~~~
\eeqn
where $\theta^+(x^2)=\theta(x^2)\theta(x_0)$ and we defined
\beq
g(a)~=~\frac{a^2}{\mid 1+\Pi((p-k)^2)\mid^2}
\eeq
The phase-space integral over the final quark-antiquark pair
has already been done.

\noindent
The dependence of $w_b$ on the coupling constant $a$ is in
\beq
g(a)~=~\sum_{n,k=0}^{\infty}\frac{(2n+k+1)!}{(2n+1)!~k!}
(-b_0^2\pi^2)^n\Bigg(b_0\log\frac{(p-k)^2}{\mu^2}\Bigg)^k a^{2n+k+2}
\eeq
whose Borel transform is given by \cite{wise}:
\beq\label{eq:borelg}
\tilde{g}(t)~=~-\frac{\sin\pi u}{\pi b_0}\left(\frac{\mu^2}{(p-k)^2}\right)^u
\eeq
To compute $\tilde{w}_b$, we substitute the expression of $\tilde{g}(t)$
in eq.~(\ref{eq:wb}) in the place of $g(a)$.

In the framework of a $1/N$ expansion, the leading log approximation (LLA)
involves the resummation of all the terms of the form:
\beq\label{eq:leadlog}
\left(\frac{1}{N}\log\frac{Q^2}{\mu^2}\right)^k,~~~~~k=0,1,2,3...,n,...
\eeq
In the evaluation of the kernel,
only those terms in $\tilde{w}_b$ which diverge
logarithmically at $u=0$ have to be considered.
This means that we need only to
keep the terms in the trace proportional to $k^2$
or to $k\cdot p$.
It is easy to see that $k\cdot p$ is of the same order as $k^2$.
The collinear singularity is indeed produced by the virtual
states of the photon with a small invariant mass, i.e.
$(p-k)^2\sim 0$, which implies $ 2 p\cdot k \sim k^2$
(see appendix A for a rigorous proof).
The terms proportional to $(k^2)^2$, $k^2 k\cdot p$, etc.
do not give rise to any logarithmic singularity
and therefore do not contribute to the evolution kernel
(they contribute to the $O(1/N)$ hard cross section).
Therefore, once a power of $k^2$ or $k\cdot p$
has been extracted, the trace
can be evaluated in the collinear limit, $k^2=0$.
This condition together with $(k+q)^2=0$, implies
\beq
k~=~xp.
\eeq

\noindent
The rate is given by:
\beq\label{eq:borelwb}
\tilde{w}_b~=~\frac{\sin\pi u ~ \mu^{2u} }{6\pi^3 b_0 N E(p)}~
\Bigg[\frac{1}{x}I_1-I_2+\frac{Q^2}{x}I_3\Bigg]~
{\rm Tr}[x\hat{p}\gamma_{\nu}(x\hat{p}+\hat{q})\gamma_{\mu}]
\eeq
where $Q^2=-q^2>0$, and the integrals $I_1,~I_2$ and $I_3$ are given by:
\beqn\nonumber
I_1&=&\int {\rm d}^4 k~\frac{\delta^+((k+q)^2)}{k^2}
                 ~\frac{\theta^+((p-k)^2)}{((p-k)^2)^{1+u}}
\\ \nonumber
I_2&=&\int {\rm d}^4 k~\frac{\delta^+((k+q)^2)}{(k^2)^2}
                 ~\frac{\theta^+((p-k)^2)}{((p-k)^2)^{1+u}} ~ 2 p\cdot k
\\ \label{eq:integr}
I_3&=&\int {\rm d}^4 k~\frac{\delta^+((k+q)^2)}{k^2}
                ~\frac{\theta^+((p-k)^2)}{((p-k)^2)^{1+u}~n\cdot (p-k)}
\eeqn
The final result is (the computation of the
integrals is described in detail in appendix A):
\beq\label{eq:decrate}
\tilde{w}_b~=~-\frac{ \sin\pi u~\mu^{2u} }{ 12\pi^2 b_0N E(p) }~
{\rm Tr}[x\hat{p}\gamma_{\nu}(x\hat{p}+\hat{q})\gamma_{\mu}]~
\frac{ x^{u+1} }{ (Q^2)^{u+1} (1-x)^{u-1} }~A(x,u)
\eeq
where
\beqn\nonumber
A(x,u)&=&
 \frac{1}{xu^2}+\frac{1}{xu}
-\frac{1}{u(u-1)}\Bigg(1+u-\frac{1}{x}\Bigg){\rm F}(1,1-u,2-u;1-x)+
\\ \nonumber
&&~~~~~~~+~2\frac{\log x}{(1-x)^2}\frac{1}{u} {\rm F}(1,-u,1-u;-x)~+~
\\
&+&\frac{4}{(1-x)^2}~{\rm LerchPhi}(-x,2,-u)~+~
\frac{2}{(1-x)^2}~{\rm h}(x,u).
\eeqn
The special functions are defined in appendix A.

Let us end this section with some comment about the decay rate
(\ref{eq:decrate}).
The single pole at $u=0$
is the collinear singularity and replaces,
loosely speaking, the pole in $\epsilon$ of usual dimensional
regularization. Notice that $A(x,u)$ has a double pole at $u=0$,
which is converted into a single pole by the factor $\sin\pi u$
coming from the transform of $g(a)$ (cf. eq.~(\ref{eq:borelg})).
The softening of the singularity occurs because
infrared singularities come from the integration over $x$
around the elastic region.
The complete box diagram involves instead the Borel transform of
\beq
\frac{a}{1+\Pi((p-k)^2)},
\eeq
given by
\beq\nonumber
\left(\frac{\mu^2}{(p-k)^2}\right)^u~~~~~~~~~~
{\rm for}~~~(p-k)^2~<~0
\nonumber\eeq
and
\beq
\left(\frac{\mu^2}{(p-k)^2}\right)^u~e^{i\pi u}~~~~~
{\rm for}~~~(p-k)^2~>~0
\eeq
In this case the factor $\sin\pi u$ is absent and there is a double pole
at $u=0$ in the final result,
coming from the product of the infrared and the collinear
singularity.

\section{The kernel}
\label{ap}

Collecting formulas (\ref{eq:lo}), (\ref{eq:zero}) and
(\ref{eq:decrate}), we have for the total rate:
\beqn
&&\tilde{w}~=~\tilde{w}_0+\tilde{w}_a+\tilde{w}_b~=
\\ \nonumber
&&~~~\frac{\pi}{2 E(p)~Q^2}~
{\rm Tr}[x\hat{p}\gamma_{\nu}(x\hat{p}+\hat{q})\gamma_{\mu}]~
q(x,Q^2/\mu^2;t)
\eeqn
where $q(x,Q^2/\mu^2;t)$ is the parton
density of a quark in the Borel plane,
\beq
q(x,Q^2/\mu^2;t)~=~\delta(1-x)\delta(t)
-\frac{ \sin\pi u ~\mu^{2u} }{ 6\pi^3 b_0 N }
 \frac{ x^{u+1} }{ (Q^2)^u~(1-x)^{u-1} }~A(x,u).~~~
\eeq
The evolution kernel is computed by taking
the derivative with respect to $\log Q^2$ of the parton density:
\beqn\nonumber
&&P_{qq}(x,Q^2/\mu^2;t)~=~Q^2\frac{\partial}{\partial Q^2}~q(x,Q^2/\mu^2;t)~=~
\\
&&~~~~~\frac{1}{6\pi^3 b_0 N}\left(\frac{\mu^2}{Q^2}\right)^u
\frac{x^{u+1}}{(1-x)^{u-1}}~u\sin \pi u~A(x,u)
\eeqn
Notice that the evolution kernel in the coupling constant space
and in the Borel space are computed in the same way, because
taking the derivative with respect to $\log Q^2$ commutes with
the Borel transform.

The factor $u\sin \pi u$ cancels the single as well as the
double pole at $u=0$ in $A(x,u)$. The evolution kernel is therefore finite
at $u=0$, as it should be for an infrared safe quantity.
Notice that the parton density itself has a simple pole at $u=0$,
coming from the collinear singularity.
The function $u\sin\pi u$ has simple zeros at the integers $u=1,2,3...,n,...$.
The singularities of the Borel transform on the real axis originate therefore
only from the term proportional to the LerchPhi function:
\beqn\nonumber
P_{qq}(x,t)&=&\frac{2}{3 \pi^3 b_0 N}
\left(\frac{\mu^2}{Q^2}\right)^u\frac{ x^{u+1} }{ (1-x)^{u+1} }~u\sin\pi u~
{\rm LerchPhi}(-x,2,-u)
\\
&+&{\rm (regular~function)}
\eeqn
There are ultraviolet renormalons corresponding to simple poles
at the integers $u=1,2,3,...,n,...$.
The leading renormalon is at $u=1$ and contributes to the
evolution kernel by a term of the form:
\beq
P_{qq}(x,t)~=~\frac{2}{3 \pi^2 b_0 N}~
\frac{\mu^2}{Q^2}~\frac{x^3}{(1-x)^2}~\frac{1}{u-1}~
+~{\rm (higher~order~renormalons)}
\eeq

\noindent
The leading non-perturbative correction coming from the
ultraviolet renormalon at $u=1$ is therefore given by:
\beq
\delta P_{qq}(x,a)~=~-\frac{2}{3\pi^2b_0^2 N}~
\frac{x^3}{(1-x)^2}~\frac{\mu^2}{Q^2}~e^{ 1/b_0a }
\eeq
where we have taken the pole residue as a measure of the ambiguity.

We can have a qualitative picture of the ambiguity in $QCD$
by changing the sign of the $\beta$-function, i.e. considering
the case $b_0<0$:
\beqn\nonumber
&&\delta P_{qq}(x,a)~=~-\frac{2}{3\pi^2b_0^2 N}~
\frac{x^3}{(1-x)^2}~\frac{\mu^2}{Q^2}~e^{ - 1/\mid b_0 \mid a(\mu^2) }~+~
{\rm (h.o.~corrections)}~=~
\\
&&~~~~~-~\frac{2}{3\pi^2b_0^2 N}~
\frac{x^3}{(1-x)^2}~\frac{ \Lambda_{QCD}^2 }{Q^2}~+~
{\rm (h.o.~corrections)}
\label{eq:final}\eeqn
where by higher order corrections we mean terms of the form
$(\Lambda_{QCD}^2/Q^2)^n$ with $n>1$.

\section{Discussion}
\label{disc}

Eq.(\ref{eq:final}) is our main result.
It says that the evolution kernel has corrections of order
$1/Q^2$.
The most interesting result is that the coefficient of the
power correction is a function of the Bjorken variable $x$, and
scales as
\beq\label{eq:scale}
\frac{x^3}{(1-x)^2}.
\eeq
The behaviour (\ref{eq:scale}) does depend on the dynamics of the model
and cannot be derived with kinematical methods.
Non-perturbative corrections grow with $x$ and
diverge as $1/(1-x)^2$  for
$x\rightarrow 1$, i.e. in the elastic region.
This is a stronger divergence than that one
of the perturbative kernel (computed up to two loops \cite{petr}), given by
$1/(1-x)$. There exists therefore a critical value
\beq
x_c~\sim~1-\frac{\Lambda_{QCD}^2}{Q^2}
\eeq
at which non-perturbative effects dominate over perturbative
dynamics and  the parton model becomes irrelevant.
There is a consequent breakdown of the factorization in the
cross section (\ref{eq:factoriz}).
We believe therefore that our analysis has not only an interest
of principle but may have also phenomenological applications,
identifying the domain of perturbative QCD.
The values of $x_c$ are quite close to one for
any reasonable value of $Q^2$. For example,
$x_c\sim 0.97$ at $Q^2=3~$GeV$^2$, while
$x_c\sim 0.999$ at $Q^2=100$~GeV$^2$.

On the contrary, non-perturbative corrections vanish as
$x^3$ for $x\rightarrow 0$. They are therefore completely
negligible with respect to the perturbative terms, which
do not vanish in the small $x$ region.
This conclusion however has to be taken with care
because we believe that QED at order $1/N$ is not completely
appropriate to describe small $x$ physics.

The qualitative features of (\ref{eq:scale}) can be understood with
the following qualitative considerations.
As $x\rightarrow 1$, the invariant mass of the hadronic
(partonic) final state and the typical virtuality of the gluon
propagator $k^2$ reduce progressively.
The effective coupling constant $\alpha_s(k^2)$
is therefore evaluated at a scale progressively closer to the pole:
$k^2\rightarrow \Lambda^2$.
The result is that non-perturbative effects increase without bound.
On the contrary, the invariant mass of the final state and
the typical virtuality of the gluon propagator diverge as
$x\rightarrow 0$.
The coupling constant is evaluated at a
scale progressively larger with respect to
the pole, and non-perturbative effects tend to zero.

\section{Conclusions}
\label{concl}

We analyzed non-perturbative effects in deep inelastic scattering
of a lepton off a hadron.
The idea is that the complete
perturbative series hiddens non-perturbative
properties of the process.
Generally speaking, these corrections turn out to be small, thereby
validating the consistency of the perturbative approach.
I am also able to explain semiquantitatively such phenomena as the
precox scaling, i.e. the scaling at quite small momentum transfer.

There is however a strong divergence of the non-perturbative terms
near the elastic region. This implies a breakdown of the parton model,
as it is conventionally understood.

We believe that our analysis can be extended to other non infrared-safe
processes, such as $\gamma-\gamma$ or hadron-hadron collisions.

\section*{Acknowledgement}

I wish to thank Z. Ligeti, D. Politzer and M. Wise for discussions.

\appendix

\section{Borel transform integrals}

In this appendix we describe the computation of the integrals
$I_1,~I_2$ and $I_3$, given in eqs.~(\ref{eq:integr}).
We also prove that the Borel variable $u$ is
a regulator of the soft divergences (i.e. infrared and collinear).

\noindent
For concreteness, let us consider $I_1$; the computation of
$I_2$ and $I_3$ is analogous. Making use of the identity
\beq
\frac{ \theta^+((p-k)^2) }{ ((p-k)^2)^{1+u} }~=~
\int_0^{\infty}\frac{d\mu^2}{(\mu^2)^{1+u}}
\delta^+((p-k)^2-\mu^2),
\eeq
the integral can be written as
\beq
I_1~=~\int_0^{\infty}\frac{d\mu^2}{(\mu^2)^{1+u}} J(\mu^2)
\eeq
where
\beq
J(\mu^2)~=~\int d^4 k ~ \frac{ \delta^+((k+q)^2)~\delta^+((p-k)^2-\mu^2) }
                           { k^2 }
\eeq
$J(\mu^2)$ is a two body phase space integral with a collinear singularity
regulated by the 'gluon mass' $\mu\neq 0$. The original integral
$I_1$ involves
an integration over $\mu^2$ with a weight function given by
$(\mu^2)^{-u-1}$. By taking  $u$ negative and large enough we can
suppress the contribution of $J(\mu^2)$ at small $\mu^2$, where the
singularity occurs, in such a way to render the integral finite.
Since (as well known) the singularities are at most powers of
logs, it is sufficient to take $u<0$.

$J(\mu^2)$ is computed by changing the loop momentum to $k'=p-k$ and
going in the reference frame in which $\vec{p}+\vec{q}=0$
and $\vec{q}$ is directed along the $z$ axis.
In this coordinate system
$p_0=\mid\vec{p}\mid=\mid\vec{q}\mid=q_s$ and
$\vec{q}\cdot\vec{k}=-\vec{p}\cdot\vec{k}=q_s k \cos\theta$.
Integrating the massless delta function, the integral becomes:
\beq
J(\mu^2)~=~\int\frac{d^3 k}{2k}
\frac{\delta^+((q_0+q_s)^2-2(q_0+q_s)k-\mu^2)}{q^2-2q_0k-2q_sk\cos\theta}
\eeq
Going to an adimensional momentum defined by $y=2k/(q_0+q_s)$ and
integrating the massive $\delta$-functions, we have the result:
\beq
J(\mu^2)~=~\frac{\pi}{ 4 q_s(q_0+q_s) }\log\frac{(1-v)r}{2-(1+v)r}
\eeq
where $v=q_0/q_s<1$ and $r=\mu^2/(q_0+q_s)^2$.

\noindent
The finite terms for $\mu^2\rightarrow 0$ generate simple poles
at $u=0$ in $I_1$, while the logarithmic term generates a double pole.

We perform now the last integration over $\mu^2$ and
express $q_0$ and $q_s$ in terms of $x$ and $Q^2$ according
to the formulas
\beq
Q^2~=~q_s^2-q_0^2,~~~~~~~~~x~=~\frac{1}{2}\Bigg(1-\frac{q_0}{q_s}\Bigg),
\eeq
whose inverse are
\beq
q_0~=~\Bigg(\frac{1}{2}-x\Bigg)\left(\frac{Q^2}{x(1-x)}\right)^{1/2}
{}~~~~~~~~q_s~=~\frac{1}{2}\left(\frac{Q^2}{x(1-x)}\right)^{1/2}
\eeq
The result is:
\beq
I_1~=~\frac{\pi}{2}~
\frac{ x^{u+2} }{  (Q^2)^{u+1} (1-x)^{u-1} }
\Bigg[-\frac{1}{x(1-x)~u^2}-\frac{1}{x~u(u-1)}{\rm F}(1,1-u,2-u;1-x)\Bigg]
\eeq
where F$=~_2{\rm F}_1(a,b,c;z)$ is the hypergeometric function,
\beq
{\rm F}(a,b,c;z)~=~1+\frac{a ~ b}{c\cdot 1}z+
            \frac{a(a+1) ~ b(b+1)}{c(c+1)\cdot 1\cdot 2}z^2+\ldots
\eeq

\noindent
The evaluation of $I_2$ is analogous to that one of $I_1$.
To simplify the angular integration,
it is convenient to write $2 k\cdot p= k^2 -\mu^2$. The integral
splits as
\beq
I_2~=~I_1~+~I_2'
\eeq
where
\beq
I_2'~=~-\int\frac{d\mu^2}{(\mu^2)^u}~K(\mu^2)
\eeq
and
\beq
K(\mu^2)~=~\int d^4 k ~ \frac{ \delta^+((k+q)^2)~\delta^+((p-k)^2-\mu^2) }
                           { (k^2)^2 }
\eeq
$K(\mu^2)$ has a power singularity of the form $1/\mu^2$ for
$\mu^2\rightarrow 0$ which, integrated together with
the factor $(\mu^2)^{-u}$, generates a pole at $u=0$:
\beq
I_2'~=~\frac{\pi}{2}\frac{ x^{u+1} }{ (Q^2)^{u+1}(1-x)^{u-1} }
      ~\Bigg[\frac{1}{xu}-\frac{1}{u-1}{\rm F}(1,1-u,2-u;1-x)\Bigg]
\eeq
The total result is:
\beqn\nonumber
I_2&=&\frac{\pi}{2}~
\frac{ x^{u+1} }{ (Q^2)^{u+1}(1-x)^{u-1} }
\Bigg[~-\frac{1}{(1-x)~u^2}
\\
&&~~~~~+\frac{1}{x~u}
      -\frac{u+1}{u(u-1)}~{\rm F}(1,1-u,2-u;1-x)~\Bigg]
\eeqn
$I_2$ has the same degree of singularity as $I_1$ at $u=0$,
given by simple and double poles.
This implies that $k^2$ and $2k\cdot p$ have to
be considered of the same order in evaluating the collinear
singularities.

For $I_3$ we have the formula:
\beqn\nonumber
I_3&=&\frac{\pi}{2}~\frac{ x^{u+2} }{ (Q^2)^{u+2}(1-x)^{u-1} }
\Bigg[~-2\frac{\log x}{(1-x)^2 } \frac{1}{u}{\rm F}(1,-u,1-u;-x)
\\
 &&~~~~~-\frac{4}{(1-x)^2}{\rm LerchPhi}(-x,2,-u)-\frac{2}{(1-x)^2}{\rm h}(x,u)
\Bigg]~~~
\eeqn
where
${\rm LerchPhi}(z,s,a)$ is the Lerch trascendent of $\Phi$, defined by:
\beq
{\rm LerchPhi}(z,s,a)~=~\sum_{n=0}^{\infty}\frac{z^n}{(a+n)^s},
\eeq
with the terms for which $a+n=0$ omitted,
and h$(x,u)$ is a function defined by the integral:
\beq
{\rm h}(x,u)~=~\int_0^1 {\rm d}y~\frac{\log[1-(1-x)y]}{y^{1+u}~[1+xy]}.
\eeq
This function has simple poles at the integers:
\beq
{\rm h}(x,u)~=~(1-x)\frac{1}{u-1}+\frac{(1-x)(1-3x)}{2}\frac{1}{u-2}
+\ldots
\eeq

\section{Planar gauge}

In this appendix we prove that collinear singularities decouple
from the final quark leg in the planar gauge (\ref{eq:sumpol}),
so that the only diagrams contributing to $P_{qq}$ are box diagrams
\cite{lip}.
In the collinear limit,
\beq\label{eq:coll}
k~=~xp,
\eeq
the polarization sum becomes
\beqn\nonumber
S_{\mu\nu}(p-k,n)&=&-g_{\mu\nu}~+~
         \frac{ p_{\mu}(xp+q)_{\nu}+p_{\nu}(xp+q)_{\mu} }{ p \cdot q}
\\
&&~~~~~~~~~~~~~+~{\rm corrections~of~order~}k^2.
\eeqn

\noindent
It is a projector orthogonal to the scattering hyperplane, i.e. the hyperplane
spanned by $p$ and $xp+q$:
\beq
p_{\mu}~S^{\mu\nu}(p-k,n)~=~0,~~~~~~~~(xp+q)_{\mu}~S^{\mu\nu}(p-k,n)~=~0.
\eeq
The trace coming from the fermion line is a tensor $T^{\mu\nu}$
lying in the scattering hyperplane in the limit (\ref{eq:coll}):
$T^{\mu\nu}~=~T^{\mu\nu}(p,xp+q)$.
Therefore, the contraction with
$S_{\mu\nu}$ vanishes:
\beq\label{eq:vanish}
T^{\mu\nu}S_{\mu\nu}~=~0~~~~~~~~{\rm at}~~~k^2~=~0.
\eeq
The diagrams with a photon radiated by the final quark contain
a single quasi-real propagator, and the integrals are of the
form:
\beq
\int\frac{{\rm d} k^2}{k^2}~S_{\mu\nu}T^{\mu\nu}
\eeq
Because of eq.~(\ref{eq:vanish}), these diagrams do not contain collinear logs.

\noindent
In the box-diagrams instead, there are two quasi-real quark propagators,
and the integrals are of the form
\beq
\int\frac{{\rm d} k^2}{(k^2)^2}~{T'}_{\mu\nu}S^{\mu\nu}.
\eeq
The terms in ${T'}_{\mu\nu}S^{\mu\nu}$ proportional to $k^2$ give rise
to the collinear log. C.V.D.

\newpage

\centerline{\bf FIGURE CAPTIONS}

\vskip .7 truecm
\begin{enumerate}
\item[Fig.1:] bubble summation in the photon propagator;
\item[Fig.2:] lowest order diagram for the scattering of a quark in an
external e.m. field;
\item[Fig.3:] diagrams of order $\alpha$ for the scattering of a quark in an
external e.m. field;
\item[Fig.4:] Diagrams of order $1/N$ for the scattering of a quark in an
external e.m. field.
\end{enumerate}


\begin{thebibliography}{99}
\bibitem{bpq}
G. Altarelli and G. Parisi, Nucl. Phys. B 126 (1977) 298;
Y. Dokshitser, Sov. Phys. JETP, 46:641 (1977);
A. Baier, V. Fadin and V. Khoze, Nucl. Phys. B 65:381 (1973).
\bibitem{books}
For an introduction see for example
{\it An introduction to quantum field theory},
G. Sterman, Cambridge University Press (1993), chapter 14.
\bibitem{politz}
A. De Rujula, H. Georgi and H. D. Politzer, Phys. Rev. D 15, 2495 (1977).
\bibitem{ope}
K. Wilson, Phys. Rev. 179 (1969) 1499;
\bibitem{ope2}
D. Gross and F. Wilczeck, Phys. Rev. D 8, 3633 (1973); ibid, D 9, 980
(1974); H. Georgi and H. D. Politzer , Phys. Rev. D 9, 416 (1974);
L. Maiani, G. Martinelli and C. Sachrajda, Nucl. Phys. B 368 (1992) 281.
M. Crisafulli, V. Gimenez, G. Martinelli et al., HEPLAT-9412049.
\bibitem{renorm}
G.'t Hooft, in {\it The Whys of subnuclear physics}, Proc. Int. School
Erice, Italy, 1977, ed. A. Zichichi (Plenum, New York, 1978);
G. Parisi, in {\it Hadron structure and lepton-hadron interactions},
Cargese 1977, ed. M. Levy et al. (Plenum, New York, 1979);
\bibitem{qedn}
D. Espiru et al., Z. Phys. C 13 (1982) 153;
A. Palanques-Mestre and P. Pascual, Comm. Math. Phys. 95 (1984) 277;
R. Coquereaux, Phys. Rev. D 23 (1981) 2276;
H. Kawai, T. Kinoshita and Y. Okamoto, Phys. Lett. B 260 (1991) 193.
\bibitem{ben}
M. Beneke, Nucl. Phys. B 405 (1993) 424.
\bibitem{wise}
A. Manohar and M. Wise, UCSD/PTH 94-11 and CALT-68-1937 preprint,
accepted for publication by Phys. Lett. B.
\bibitem{petr}
G. Curci, W. Furmansky and R. Petronzio, Nucl. Phys. B 175 (1980) 27.
\bibitem{lip}
L. Lipatov, Sov. J. Nucl. Phys., 20:94, 1975.
\end{thebibliography}
\end{document}